# Semantic Communications in Networked Systems:
# A Data Significance Perspective

Elif Uysal, Onur Kaya, Anthony Ephremides, James Gross, Marian Codreanu, Petar Popovski, Mohamad Assaad, Gianluigi Liva, Andrea Munari, Beatriz Soret, Touraj Soleymani and Karl Henrik Johansson

## ACKNOWLEDGEMENT

This paper was written based on two European project proposals by the authors. These proposals were prepared in cooperation with Vangelis Angelakis, Nikolaos Pappas, Adam Molin, Mert Ankaralı, Elif T. Ceran, Dina Bousdar, Xavier Vilajosana, Vahid Mamduhi and Stephan Sand. The authors would like to thank all partners for their collaboration, and Sajjad Baghaee and Federico Chiariotti for their help with the figures.

## ABSTRACT

We present our vision for a departure from the established way of architecting and assessing communication networks, by incorporating the semantics of information, defined, not necessarily as the meaning of the messages, but as their significance, possibly within a real-time constraint, relative to the purpose of the data exchange. We argue that research efforts must focus on laying the theoretical foundations of a redesign of the entire process of information generation, transmission and usage for networked systems in unison by developing (1) advanced semantic metrics for communications and control systems; (2) an optimal sampling theory combining signal sparsity and timeliness, for real-time prediction/reconstruction/control under communication constraints and delays; (3) temporally effective compressed sensing techniques for decision making and inference directly in the compressed domain; (4) semantic-aware data generation, channel coding, packetization, feedback, multiple and random access schemes that reduce the volume of data and the energy consumption, increasing the number of supportable devices. This paradigm shift targets jointly optimal information gathering, information dissemination, and decision-making policies in networked systems.

## INTRODUCTION

The cornerstone of the evolution of communication systems since the 1940's has been the paradigm that humans choose the data, while the network ensures its correct, and to some extent timely, delivery. Today, communication networks are evolving from being highly optimized to deliver data to humans, towards modes that cater for autonomous interactions among machines and things. This is aligned with recent research initiatives on 6G and other beyond-5G systems, as well as the projected growth of the Internet, which is expected to be dominated, in terms of both the number of nodes and economic capacity, by *networked applications* generating machine-type data. These real-time systems are driven fundamentally by the necessity to automate decisions in a sense-compute-actuate cycle. Scalable support for such systems depends on what we refer to as "*semantic communication*": the provisioning of the right and significant piece of information to the right point of computation (or actuation) at the right point in time; a notion that carries elements from Level B and C communication problems proposed by Weaver [1]. However, to date the theoretical basis for semantic communication is incomplete, because **in classical data communication "the right or significant piece of information" is not defined, nor is the process of information generation factored into the communication process**: The goal is to reliably transmit a given data stream in its entirety as fast as possible. This approach can be dramatically wasteful for applications that require only selected parts of the information in the data stream. Consider the following two examples:

**1. Sparse sampling in remote estimation.** In smart manufacturing systems, the states of plants often need to be transmitted to remote centers for estimation. However, communication between plants and centers can involve random delay. In this case, it has been shown that the estimation performance, measured by mean square error (MSE), can be improved by orders of magnitude using *process-aware sparse sampling*, as opposed to uniform sampling [2]. This promises significant reduction in sampling rate and energy expenditure on the device side, critically important for low-power/energy harvesting sensors (e.g., IoT nodes). It furthermore reduces network resource consumption.

**2. Timely consensus among autonomous vehicles.** In intelligent transportation systems, negotiation for timely consensus among autonomous vehicles about intended maneuvers is essential to avoiding collisions. This is vital in unforeseen situations, e.g. sudden appearance of pedestrians. *Semantic-aware transmission*, which respects the time-dependent value of the messages, can enable the network to prioritize the information flow efficiently while meeting safety demands.

Other applications relying on real-time decision making, such as smart grids and networked control, also demand a restructuring of the communication process. In all these cases overprovisioning stemming from the generation and transmission of "raw" data streams creates enormous data volume, leading to network bottlenecks. **These bottlenecks, if left unresolved, will severely limit the growth and utility of networked systems.** To date, there is a major knowledge gap between existing data communication paradigms and the needs of proliferating real-time systems in different domains, whose growth is not supportable by these paradigms.

Resolving these bottlenecks requires a fundamental transformation of established practice in communication system design and a complete re-formulation of data **generation** and **transfer.**

In this paper, we present a vision of how this gap may be closed by moving toward a **semantic communication architecture (Fig.1)**. We envision an end-to-end architecture that enables the **generation of just the right amount of data and transmission of the right content to the right place at the right time**. This will be achieved through a **redesign of the entire process of information generation, transmission and usage in unison**, by incorporating *the semantics[1] of information,* defined, not necessarily as the meaning of the messages, but as their *significance*, possibly within a real-time constraint, relative to the purpose of the data exchange. As Fig.1 shows, semantic requirements translate to comparative metrics in relation to other flows, at different stages in the communication architecture, for instance in resolving contention in multiple access.

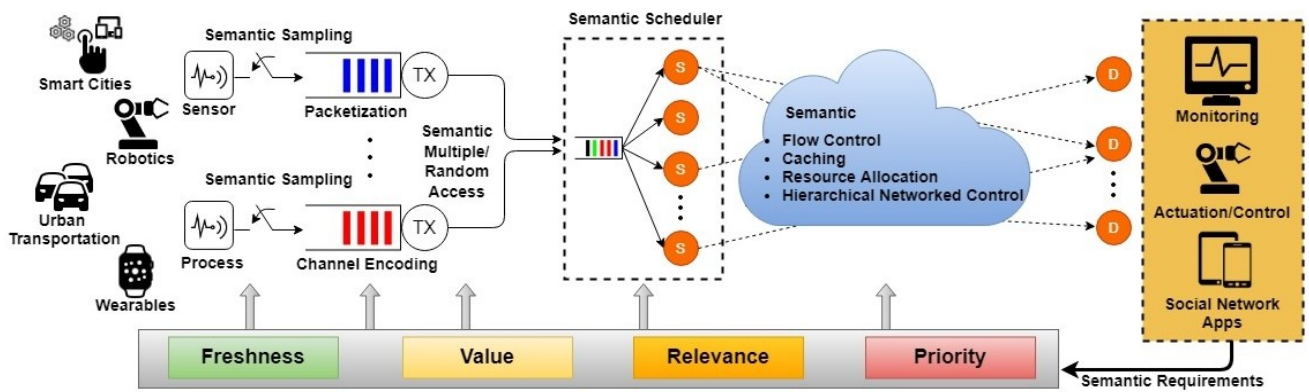

Fig.1. Components of the envisioned end-to-end **Semantic Communication architecture.**

### A SEMANTIC-AWARE END-TO-END ARCHITECTURE FOR INFORMATION FLOW

The envisioned architecture in Fig.1 must be built on new theoretical results incorporating semantics-based metrics, as exemplified in the following section, into the communication process. This vision goes beyond the current technological paradigms in the following aspects:

- Most existing and developing communication paradigms (including 5G) are based on the assumption of uncontrolled arrival of exogenous traffic to the communication system, which sets design boundaries to protocols. It is essential to *relax the assumption of exogenous data arrivals* and explicitly enable non-uniform semantic-aware sampling within the communication system design.
- Current paradigms for channel encoding, multiuser scheduling, channel access, and flow control are essentially agnostic of semantic properties of the data, which leads to inefficiency. We argue that link, transport and application layer principles must be developed in concert in order to fulfill semantic-related targets under scarce network resources and energy.
- Current NCSs have to live on top of classical communication networks which have been optimized for data streaming or file transfer. We envision new communication protocol principles tailored to the **optimization of the information flow in control systems**.
- While IPv4, IPv6 and the latest releases of the 3GPP enable prioritizing traffic based labels such as Type of Service and Traffic Class, and network controllers add meaning to these fields by defining priorities in their traffic flows, these approaches have been implemented with the classical view of the generation and transmission of data, and are thus limited in tackling the upcoming growth of network congestion.

---

[1] The word *semantics* etymologically roots from the ancient Greek word "Semantikos" ("significant") and has later evolved, in the context of languages, to "meaning". Our use of the term is more correlated to the original connotation, applied to the context of machine-type communication.

We next elaborate on some semantic measures which will serve as objectives in our envisioned semantics-aware design approach for communication systems and networks.

# EXAMPLES OF SEMANTIC MEASURES FOR THE ENVISIONED ARCHITECTURE

The semantic and effectiveness problems were initially recognized in [1], predominantly for the case of human communication. in which context they refer to the conveying of desired meaning to, and the effective influencing of desired conduct at the receiver, respectively. Here, we discuss attributes of data such as timeliness, priority, and value capturing its "significance" required for the "effective" operation of the receiver, in machine type communications (MTC), particularly with respect to real-time decision scenarios. In this section, we further illustrate our perspective, by describing a set of semantic measures. We note that there are other emerging notions of semantics in communication, such as [3,4,5] that are closer to the definition of semantic as "meaning"; and that for MTCs with human-in-the- loop, the relation between meaning and significance should also be investigated.

## FRESHNESS

A quantitative and analysable surrogate for semantics in some applications (e.g., situational awareness, location tracking, control) is the **Age of Information (AoI)**: a measure of **freshness** for an information flow, defined as the time that has elapsed since the newest sample of this flow available at the destination was generated at the source. Let $u(t)$ be the timestamp of the newest packet that the destination has received by time $t$. The age of this flow at $t$ is $\Delta(t) = t - u(t)$. Recent literature [6] showed that performance in many applications can be optimized through controlling age, a penalty function $g(\Delta(t))$ of age, or other emerging metrics related to age, such as Query-AoI (QAoI) [7] and Age of Incorrect Information (AoII) [8].

AoI is a composite measure that depends on both **the sampling pattern** used to generate data from the source signal, and **the delay** these samples encounter in the network. Hence, this measure goes against state-of-the-art network design principles that handle these aspects separately. Standard transport protocols (e.g.,TCP and UDP) do not explicitly support freshness, which can lead to extreme inefficiency [9].

The current 5G Ultra-Reliable Low-Latency Communication (URLLC) paradigm mandates a departure from utility-based network design that relies on average quantities, to deliver strict guarantees on *latency*, defined as the delay from the transmission of a data packet to its successful decoding at the receiver. However, **controlling latency is not sufficient**: many machine-type applications require freshness, which can be measured by the Peak AoI (PAoI). Consider the MQTT transport protocol widely used in industrial applications: Even if the 3GPP release-16 URLLC goal of 0.5-1.0 ms one-way latency with 99.9999 percent transmission reliability is met, this does not guarantee freshness: e.g., with a random sampling period of 1 ms, PAoI will reach 2 ms. Lowering the PAoI to 1.1 ms would require the sampling period to be lowered to 0.1 ms, whereby *the sampling rate is tailored to the latency provided by the network*. Noticing this distinction between latency and freshness is critical in networked control systems (NCSs), considering the smart manufacturing market, expected to grow to €0.9 trillion by 2030, to which wireless connections will contribute 72% (According to a news article published May 12, 2020 by Ericsson.) .

## RELEVANCE

Consider measurements of a process sent to a monitor for remote estimation or tracking, over a channel with random delay. A dramatic example for the sub-optimality of separate handling of sampling and transmission is the several-fold increase in reconstruction error suffered by state-of-the-art uniform sampling as opposed to **age-aware sampling** (Fig.2, red curve). A more advanced semantic criterion for sampling (beyond age) is the **amount of change in the process** since the previous sample, which measures the **relevance** of this sample for the computation at the monitor. Process-aware non-uniform sampling can match the MSE achieved by uniform sampling by sending only a fraction of the samples (Fig.2, blue curve). Therefore, rethinking sampling is crucial to future system designs, especially those involving low-power devices. Recently, the semantic-aware sampling problem has revealed connections with the concepts of Rényi entropy, Gallager's reliability function, and anytime capacity.

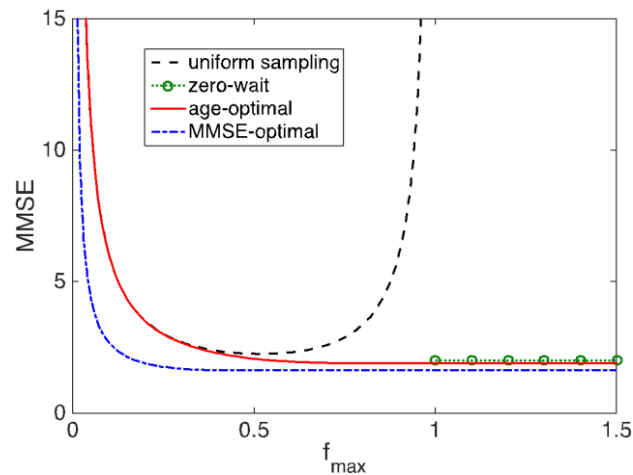

Fig.2. [2] Estimation error with respect to maximum allowed sampling rate, *when packets are subject to* IID *unit mean* exponential*ly distributed* delay in *the channel*. MMSE-optimal *and age-aware sampling* can be arbitrarily better than uniform sampling, which sets the sample rate to $f_{max}$.

## VALUE

The semantic attribute, relevance, defined above considers not only the ageing of the source sample but also the change in the source output. Yet, this may be insufficient for certain NCSs that require a more advanced metric that also takes into account the *value* of the next source sample to the point of computation. The **Value of Information** (VoI) can be quantified as the difference between the benefit of having this sample and the cost of its transmission. Transmission policies based on VoI have the potential to greatly reduce the data traffic to guarantee a given level of control performance [10] (Fig.3).

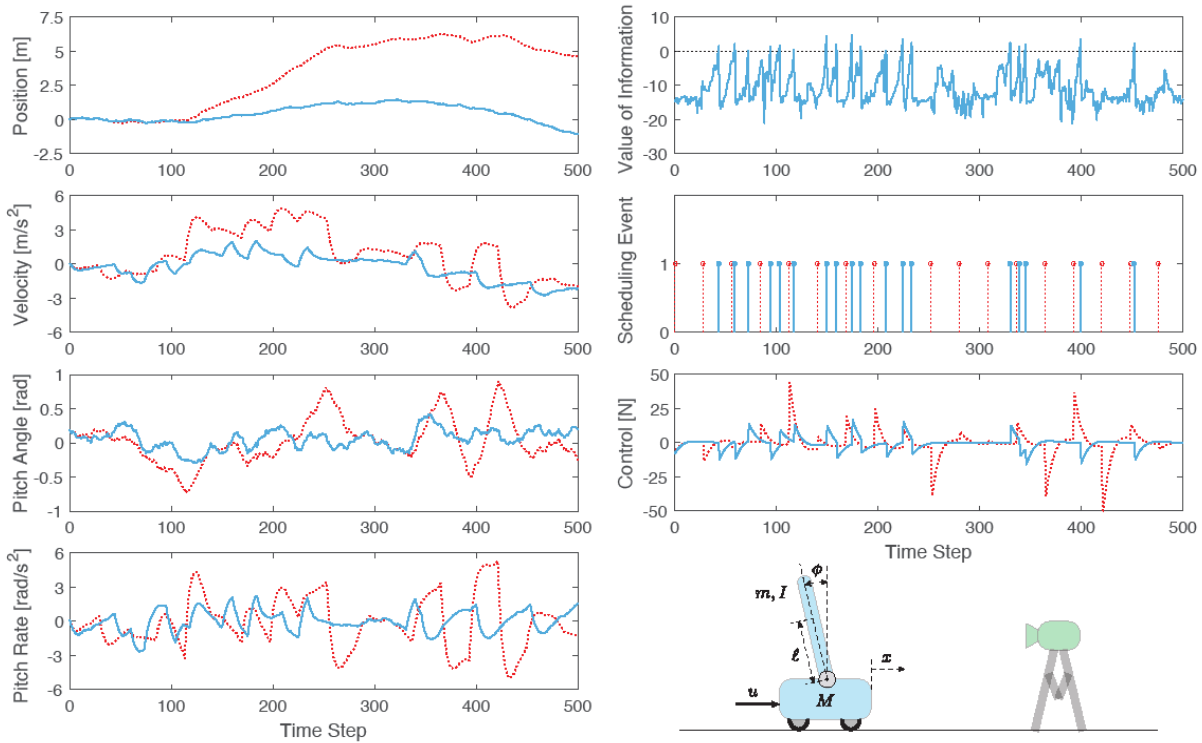

Fig.3. Feedback control of an inverted pendulum. Sensory information is transmitted to the controller based on a VoI policy [10]. The trajectories of VoI, scheduling events, and control are shown on the right, those of position, velocity, pitch angle, and pitch rate are shown on the left, in blue. In this experiment, the VoI became positive only

18 times out of 500 steps, each resulting in a message transmission to the controller. The corresponding trajectories under a periodic policy with the same number of transmissions are plotted in red.

## TOWARD NEW PERFORMANCE METRICS TO EMPOWER THE SEMANTIC-AWARE COMMUNICATION ARCHITECTURE

Each of the semantic attributes listed above are components of our definition of semantic communications: *Freshness* is related to "the right time" to send a new update. *Relevance* is related to generating "the right piece of information" through sampling. *Value,* in addition, is related to provisioning the "right piece of information", in a timely manner, to the "right point of computation", especially in the context of information flow in cyber-physical and hierarchical control systems. Burgeoning research efforts have been investigating semantic metrics that combine these attributes to varying degrees by generalizing the notions of AoI, VoI, etc. as they relate to accurate prediction or control, which vary depending on the dynamics of the underlying system, the autocorrelation function of the data, and whether the data is genuine or not.

One of these new metrics is QAoI [7, 11], which combines the semantic attributes *age* and *value* by considering a pull-based system in which the information is useful at the receiver only at certain query instants. The information freshness at query instants, can be greatly improved by scheduling transmissions according to a QAoI optimized strategy as opposed to an AoI optimized one [7, 11]. The standard communication model is essentially a push-based model, in which the sender is assumed to always pick a message that is relevant and the receiver is ready to receive it. In many practical systems, such as LEO satellites that collect information from IoT devices or, it is the receiver that drives a pull-based communication model.

Information aging and accuracy are amalgamated together in a new semantic metric called AoII [8], which combines the semantic attributes *freshness* and *relevance*: roughly, if the sampled phenomenon has not changed significantly since the previous update, the AoII stays flat even though AoI rises. Fig.4(a) demonstrates an example of the application of this metric in video streaming: an AoII-based transmission policy reduces the distortion with respect to AoI-based policy which is already superior to the conventional error-based policy. Fig.4(b) illustrates a sharper significance of using AoII in an industrial setting.

Sparsity and information aging can be combined with the delay statistics of different network paths to derive radically new methods of semantic-aware causal signal reconstruction from sub-Nyquist randomly sampled signals.

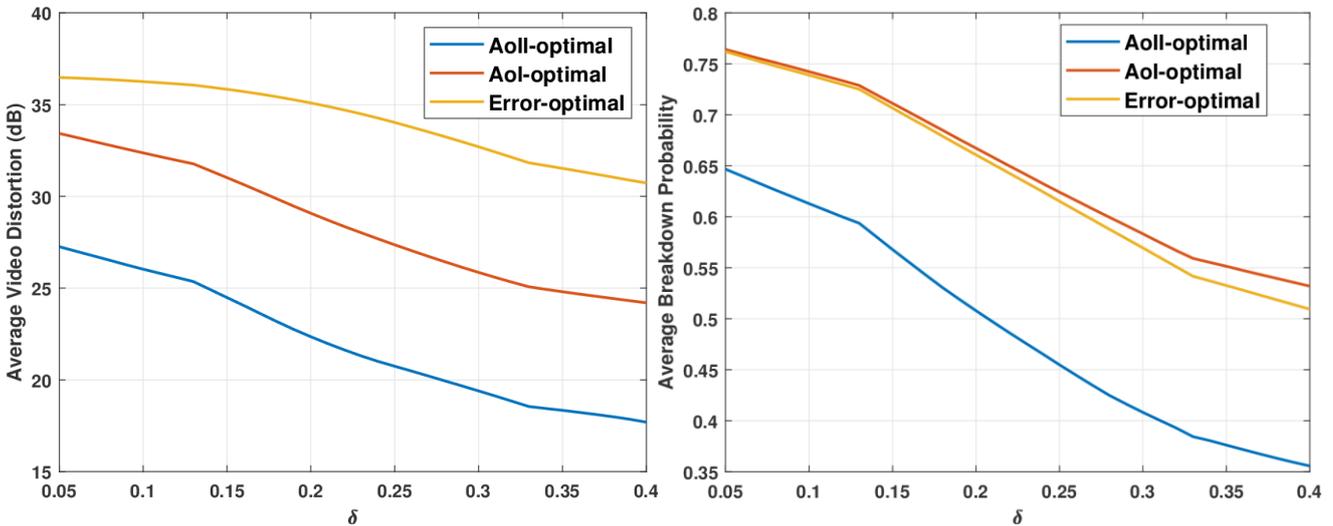

Fig.4. (a) Reduction in Video Distortion, (b) reduction in machine breakdown probability in an industrial setting, with AoII-optimal sampling compared to AoI-optimal and the conventional error-optimal scheme. $\delta$ represents maximum cost allowed for updates [8].

Another semantic attribute, for comparison between information belonging to different data flows, is priority. While priority has been widely used in network design as an externally dictated constraint or the solution of a resource allocation problem, it can alternatively be derived from the data as a semantic property, in many scenarios (e.g., real-time consensus in urban transportation systems including high speed trains, vehicles, vulnerable road users; satellite asset tracking).

Next, we explore the challenges related to the envisioned architecture in Fig.1.

## CHALLENGE #1: SAMPLING, SOURCE CODING AND COMPRESSION

Virtually all state-of-the-art information processing schemes follow the same "sample-then-compress" structure: the input signal is first acquired at the desired resolution and then compressed using computation-intensive signal processing techniques that remove the redundancy. Thus, a large amount of raw data may be collected during the acquisition stage just to be thrown away at the compression stage or it ends up as unnecessary traffic injected into the network, leading to poor scalability. In sharp contrast, a *semantic compressed sensing* approach would be a computationally lighter method which collects only the minimum amount of data needed to reconstruct the signal of interest *at the desired resolution, as orchestrated by the application requesting the data*.

Semantic compressed sensing would perform certain signal processing operations (e.g., detection, classification, learning) directly in the "compressed domain", without complete signal reconstruction. This leads to the ambitious goal of developing **an optimal sampling theory** combining signal sparsity and aging/semantics, for real-time prediction/reconstruction **under communication constraints and delays.**

This vision on sensing diverges significantly from the conventional compressed sensing framework [12], where the goal is to reconstruct a signal based on a set of measurements of equal importance, in that: (i) the collected samples/measurements are augmented with **semantic attributes** that can encode for example the reliability and/or accuracy of each collected sample (noisy, outdated or unreliable samples can be more coarsely quantized or even discarded) and (ii) it utilizes, jointly, aspects of sparsity and information aging to reconstruct **causally** accurate approximations of the monitored processes (or to estimate certain functions/properties/parameters of them) from streams of **sporadic measurements** acquired with an average sampling rate below the Nyquist limit. As these aspects have so far been studied separately, our vision is a common mathematical framework that handles them in unison. This vision goes beyond the traditional notions of source and channel coding, including joint source-channel coding (JSCC), as it respects the time constraint in the encoding and reconstruction. For example, information updating in autonomous vehicles requires a real-time compression that enables the receiver to reconstruct the highest priority source messages in a timely manner under limited network resources. This means that *there is no time for traditional source coding* that focuses on minimizing the codeword length in order to approach the Shannon entropy of the source. Solving this problem **cannot rely on classical JSCC**, which aims to entirely reconstruct an information stream, by combining the channel feedback with the source entropy.

Communication systems have converged to achieve the fundamental limits of classical data communication, where the key performance indicators (KPIs) are **throughput** and **latency**. For example, one pillar of 5G is URLLC that is focused on latency, while another pillar, eMBB, maximizes the throughput. However, it has been shown that **maximal throughput or minimum delay are neither necessary nor sufficient** for optimal operation in rapidly evolving networked applications based on status updates and remote computations. The reservation of URLLC resources to support overly stringent latency requirements limits the resources that can be allocated to broadband traffic. However, such requirements may often be artificially created by the **suboptimality of the exogenous data arrival assumption**. This can be alleviated, if, *only data that is truly vital for the application is generated, respecting the state of the network, through the methodological innovations with the definition of semantic metrics*. Relaxing the exogenous arrival assumption will enlarge the design space for channel encoding, multiuser scheduling, channel access, networking and flow control for MTC.

## CHALLENGE #2: PACKETIZATION, SCHEDULING, RESOURCE ALLOCATION

In order to achieve optimal trade-offs between semantic metrics and **energy consumption** in the network, the structure of **feedback, packetization, scheduling and resource allocation** must be revised. This requires abandoning simplistic assumptions (e.g. devices always have fresh information, packet transmission consumes unit energy) and simple metrics (AoI) and potentially exploiting machine learning methods.

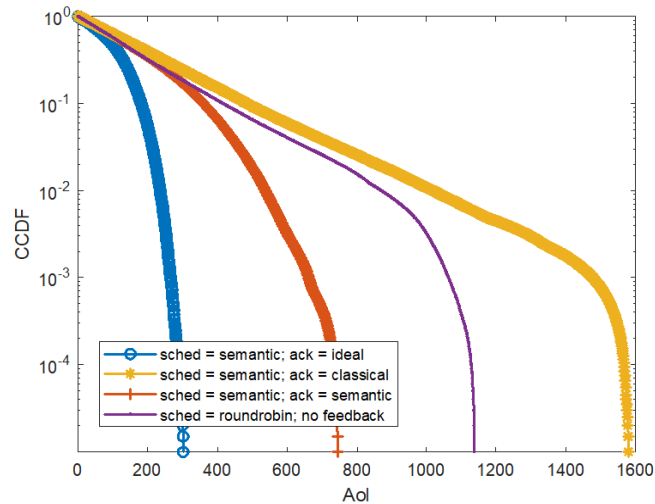

Fig.5. Impact of semantic-aware feedback and scheduling in a multiuser downlink.

Fig.5. demonstrates the importance of incorporating semantic design of feedback, in a packet scheduling problem at a Base Station (BS). When the portion of the frame allocated for feedback is constrained so that the BS is able to give feedback to only selected users in each frame, semantic-aware feedback -- rather than the classical packet-based feedback where users are given equal opportunity-- achieves significantly better performance.

Many applications within the vision of this paper generate short packets, a significant portion of which is spent on **metadata** as well as auxiliary operations, such as packet detection. Therefore, the packet structure (data vs metadata) must be optimized, and the role of feedback and link layer retransmissions in realistic channel conditions and bit error rates achievable under non-asymptotic block lengths must be investigated, with the goal of meeting semantic objectives.

It is also crucial to develop Semantic-aware scheduling and resource allocation policies that factor in **energy availability,** a major issue in evolving networks. Contrary to widespread intuition that getting fresher data requires a larger sampling rate, it has been shown [13], for example, that one can not only save energy by sending updates at 30% reduced rate, but even achieve improved freshness (specifically, 6% lower AoI) in the process. This opens up fascinating possibilities for novel low power systems. The energy implication of such reduction in the number of transmissions for billions of terminals will be enormous.

## CHALLENGE #3: SEMANTIC-AWARE MULTIPLE AND RANDOM ACCESS

Advanced access techniques under consideration for future 3GPP releases of cellular communications, such as Non-Orthogonal Multiple Access (NOMA), focus on maximizing the number of users or throughput. Whilst NOMA provides spectral efficiency enhancement, this does not directly translate into better performance in terms of freshness [14] or other semantic attributes. In addition to scheduled access, establishing semantic principles for random access (CSMA, CSA, E-SSA, frameless ALOHA) for status update flows of next-generation massive machine type communications (mMTC), IoT, and applications remains a fundamental problem, calling for novel protocols that incorporate sampling and data generation discussed in previous sections. Preliminary works have already shown great improvement through age based random access (Table I). Rethinking random or scheduled access with respect to broader semantic metrics remains open.

|  | Threshold ALOHA | SAT | MiSTA |
|---|---|---|---|
| **Spect. Efficiency improvement** | none | none | 43% |
| **Average age reduction** | 48% | 50% | 63% |
| **Energy saving** | 10% | 64% in Tx, ~100% in Rx mode | ~40% |

Table I: Improvement in spectral efficiency, freshness and energy consumption with respect to slotted ALOHA, achieved by several age-aware random access protocols proposed in recent literature ([15], and references therein).

# CHALLENGE #4: OPERATION AT THE TRANSPORT LAYER AND ABOVE

The design of flow control algorithms is inevitably connected with the generation of samples, which dictates the input flow to the system. Sampling, flow control and lower-layer retransmissions complement each other in the model in Fig.1. Novel flow control mechanisms (e.g., the Age Control Protocol by Shreedar et al.) and application-layer queuing/service disciplines circumventing suboptimal operation of existing transport-layer protocols can tremendously improve the use of network resources.

## APPLICATIONS
### NETWORKED CONTROL SYSTEMS

State-of-the-art communication technologies are agnostic to control objectives. However, separate optimization of communication networks and control systems can be highly suboptimal, as it narrows down the solution space of the problems in both areas. The control system designer, not having access to flexibility in the network design, might impose unnecessarily strict constraints that the network design has to then accommodate. A breakthrough toward enabling massive scale NCSs is not possible using techniques from each field separately, as each has already approached its limits. This calls for unifying techniques from the aforementioned disciplines under the umbrella of semantics of information, re-designing information generation, transmission, transport, reconstruction techniques to directly optimize the performance of the application that uses this information.

A first step toward this is to characterize *trade-off regions* between control costs and information costs, and use these as guidelines to design policies for scheduling data packets (samples) (Fig.3), respecting processing and communication delays in observation/command channels. Moreover, it is important to study the joint design problem in *hierarchical networked control systems* and to determine *optimal decision-making points* in the hierarchy under various information structures. The high-level decision makers, compared to low-level ones, have more information about the whole system, but communicating with them is costlier and subject to larger delays. Already, there are emerging protocols that cater better to NCSs through the use of semantic metrics . However, overall, further aggregation towards our vision in Fig.1 is required when scaling such approaches to larger networked systems.

### SMART CITIES

Cities are evolving towards complex digital-human ecosystems in which different ICT and physical technologies and human beings coexist. Semantic-aware ICTs can facilitate this human-digital ecosystem, connecting machines, communities, physical things, processes, and content, in which the limitations of current technologies will be overcome through a data/information centric perspective.

For instance, in urban mobility scenarios, semantic-aware RAN and networking functions can accommodate the cooperation of V2X and smart city infrastructures according to the criticality of the situation. One way to determine the criticality is through the VoI concept discussed above. Actuation and sensing systems in a city can react to emergency situations based on an evaluation of VoI, ceasing normal operation and prioritizing the movement of authorities (police, ambulance), turning semaphores green/red to delimit areas, etc. Another application is health related wearables or mobile apps that report vitals and location data. This is a growing sector in home care and infectious disease control.

Availability of reliable, low-latency wireless networks, capable of connecting several subsystems (automated vehicles, trains/subways, transport infrastructure, parking areas, etc.) is a key enabler of smart mobility. It is essential to leverage semantic measures to organize the data traffic in accordance with the limited available wireless spectrum resources, to reach safety critical timely consensus.

### mMTC/IoT SYSTEMS

By 2025 the number of connected devices is expected to approach 100 billion, providing services for, e.g., smart agriculture, industrial and environmental monitoring, metering, fleet and asset tracking. In such applications, devices may be located in remote areas, with satellite connectivity playing a key role. Spectrum is scarce and expensive, implying that massive populations of terminals will be served over narrow channels (e.g., the recently-introduced VHF Data Exchange System (VDES) for ship tracking operates in a 50 KHz band). This fundamental challenge can only be met by rethinking the communication paradigm, leveraging semantic properties of monitored processes to transmit data which is truly relevant, saving energy and enabling effective bandwidth sharing.

Semantic-driven design is already part of some IoT systems: e.g., VDES tunes the update rate to vessel location or speed, based on the notion that sudden changes in routes are more critical in busy harbour areas. Value, priority, and related semantic metrics, are thus expected to become cornerstones of future IoT systems.

## CONCLUSIONS

Realizing the vision presented in this paper calls for developments at the intersection of signal processing, communication, networking, and control. Each of these fields is vastly matured in its own right, yet they separately deal with different aspects of information: signal processing deals with generation, compression, processing and reconstruction of data; communication theory deals with its reliable transmission and detection at bit or symbol level, networking deals with its transport and delivery, and control theory deals with it as a measured state to be controlled or input to an action decision. Cross-fertilizing the formulations of these disciplines is nontrivial. Yet, recent research applying age penalties, delay violation probabilities, and other instances of semantics in certain sub-problems *handling one or two aspects or layers at a time* of the big picture in Fig.1, suggest that the envisioned clean-slate, holistic approach to sampling, transmission and the purposeful integration of information in the application can lead to substantial gains, with far reaching potential.

For example, semantic-aware causal signal reconstruction from sporadic measurements will not only advance the state-of-the-art in signal processing, but also permit intelligent inference and enable automated operation of devices that perform advanced applications, such as Industry 4.0, autonomous robots, environmental monitoring.

The proposed ideas represent a *radical departure* from the well-established way of assessing communication networks by means of throughput and delay, which challenges the current thinking about communication systems, and the established methodology for their design. This departure relies on an extensive cross-layer optimization, a methodology that has not always been unconditionally accepted by the mainstream industry and the networking community, as it reduces the transparency of layer 1-4 protocols and equipment (transceivers, switches) to the type of data. Yet, since many of the outlined innovations (e.g. sparse sampling) take place in Layers 5 and above, an impact can readily be made through shaping input traffic before putting it out on the network, being transparent to layers 3 and perhaps 4, which is a significant technical flexibility.